\documentclass[11pt]{article}

\pdfoutput=1

\usepackage{amsmath}
\usepackage{amsthm}
\usepackage{amsfonts,bm}
\usepackage{a4}
\usepackage{graphicx}
\usepackage[utf8]{inputenc}
\usepackage[english]{babel}

\usepackage{authblk}

\usepackage{enumerate}
\usepackage{enumitem}
\usepackage{float}
\usepackage{newcent}
\usepackage{mathrsfs}

\usepackage[bottom=2.5cm]{geometry}
\usepackage{cancel}
\usepackage{listings,color,verbatim}
\usepackage{fancyvrb}
\usepackage{xcolor}
\usepackage{pdfpages}

\usepackage{framed,color}
\definecolor{shadecolor}{gray}{0.94}

   {\snugshade\verbatim}%
   {\endverbatim\endsnugshade}

\newcommand\mylim[2]{%
    \begin{array}[t]{@{}l@{}}
     \lim #1 \\[-2ex] \scriptstyle \hspace{-0.6ex} #2
    \end{array}
}

\usepackage{multirow}
\usepackage{amstext}
\usepackage{array}
\newcolumntype{C}{>{$}c<{$}}

\makeatletter
\renewcommand{\maketitle}{\bgroup\setlength{\parindent}{0pt}
\begin{center}

\end{center}

\begin{flushleft}
\LARGE \textbf{\@title} \\
\vspace{0.5cm}
  \@author
\end{flushleft}\egroup
}
\makeatother

\title{Reflection on modern methods: competing risks \textit{versus} multi-state models.}

\author[1,2*]{\large Fran Llopis-Cardona}
\author[3]{Carmen Armero}
\author[1,2]{Gabriel Sanfélix-Gimeno}
\affil[1]{\small Health Services Research Unit, Foundation for the Promotion of Health
and Biomedical Research of Valencia Region (FISABIO), Valencia, Spain.}
\affil[2]{\small Red de Investigación en Servicios de Salud en Enfermedades Crónicas (REDISSEC), Valencia, Spain.}
\affil[3]{\small Department of Statistics and Operations Research. Universitat de València, Spain.}
\affil[*]{Corresponding author. E-mail: \texttt{llopis\_fracar@gva.es}}
\date{} 

\begin{document}
\maketitle
\vspace{0.5cm}

\begin{abstract}
Survival competing risks models are very useful for studying  the incidence of   diseases whose occurrence competes
 with other possible diseases or health conditions.  These models perform properly when working with terminal events, such as death, that
   imply the conclusion of the corresponding study. But they do not allow the treatment of scenarios with non-terminal competing events that may occur sequentially.
Multi-state models are complex survival models. They  focus  on  pathways defined by the  temporal and sequential occurrence of
numerous events of interest and thus they are suitable for   connecting   competing non-terminal events  as well as to manage other survival scenarios with higher complexity.
We discuss competing risks within the framework of  multi-state models and clarify the usefulness
 of both models for analysing epidemiological data. We highlight
  the power of multi-state models  through a real-world study of recurrent hip fracture from Bayesian inferential methodology.
\end{abstract}


\vspace{0.5cm}

\noindent
\fbox{\parbox{\textwidth}{

\textbf{Key messages}
\begin{itemize}[leftmargin=*]
\item We define the competing risks model as a particularization of the illness-death model, in which no transition from illness to death is allowed.

\item Both formulated models provide almost identical estimations of the cumulative incidences for the competing events.

\item We propose multi-state models as a preferable tool to estimate those incidences as they also provide evidence regarding transitions between the competing events, in particular from illness to death.
\end{itemize}
}}

\newpage
\section{Survival and event history analysis}

Epidemiology deals with the incidence and evolution of diseases in populations. Although epidemiological studies can be conducted with different objectives and perspectives, time in these projects is an important,  and often essential, element.  Survival and event history analysis also address  time and constitute   important methodological tools for the treatment of epidemiological data.  They focus  on time to the occurrence of one or several events of interest  which, in epidemiological research, are usually associated with death, the cure of a disease,  a  positive and/or negative progression of a disease, the appearance of adverse effects, etc. as well as with the possible trajectories determined by   the occurrence of different events in individuals of the target population.

Competing risk and multi-state models are particular survival models developed for the joint statistical treatment of more than one event of interest (Putter \textit{et al.}, 2007).
The first one deals with survival times  in the presence of competing events where the occurrence of a particular event  prevents the occurrence of  all remaining events. Competing risks models are very useful
 in medical and epidemiological research  and have been widely used to assess rates and risks  (Lou \textit{et al.}, 2009; Andersen \textit{et al.}, 2012; Wei \textit{et al.}, 2018;  Schuster \textit{et al.}, 2020).

There are different approaches for studying competing risk models. The two more popular ones are the  \textit{so-called}  multivariate
time to failure model,  which was the first model  proposed (Gail, 1975) in the survival   literature,  and the cause-specific hazards model subsequently introduced
by Gaynor (1993). We will focus on the latter  because it does not have the problems of lack of identifiability of the former and  connects naturally and intuitively with the multi-state framework.  They allow to estimate the cumulative incidence of those events, i.e. the proportion of individuals who would fail at any particular time,  by means of modelling the hazard rate associated to each event.

Multi-state models are complex models that focus  on  the complete event history. They are a class of stochastic models that enable
individuals to move between different states over time. These states generally  represent different conditions of illness and/or health. Relevant survival times are times between states which, from a methodological perspective, can be analyzed by means of stochastic processes and  survival procedures (Andersen and Keiding, 2002). Despite their enormous usefulness, multi-state models are not yet very popular in the world of epidemiology (Commenges \textit{et al.}, 1999;  Beyersmann \textit{et al.}, 2011; Hill \textit{et al.}, 2021).

Our aim in this paper  is to establish connections between competing risks  and multi-state models that allow us to clarify the usefulness
 of both models for analysing data from epidemiological settings as well as  to deepen  their analogies and differences. In particular, we want to emphasise the potential of multi-state models to study epidemiological processes that require the analysis of different health conditions whose occurrence may be sequential in time and always  in environments subject to uncertainty.

 This paper is structured as follows:  Section 2 reviews competing risks and multi-state models. In particular, Subsection 2.1 discusses the cause specific-hazard approach to competing risk models, and Subsection 2.2 contains the main elements of multi-state models  and describes in detail the \textit{so-called} illness-death model,   a particular type of multi-state models with three
states, two of them transient  and one terminal. Subsection 2.3 examines the competing risk models in the framework of multi-state models.   Section 3 presents a study based on data from the PREV2FO cohort, which comprises patients aged 65 years and older who were discharged after a hospitalization for osteoporotic hip fracture between 2008 and 2015.  We use a competing risks model and a multi-state model to analyse the follow-up of these patients over time with special interest in the occurrence of a subsequent hip fracture or death. Both models were analyzed via Bayesian statistics. This approach allows to obtain posterior  distributions of the different quantities of interest such as the cumulative incidence of refracture or death in the competing risks model, or the proportion of people who die or remain alive after a refracture in the case of the multi-state models. The paper ends with some concluding remarks.

\section{Competing risks \emph{versus} illness-death survival models}

\subsection{Competing risks models}

As stated before, competing risks models are  survival statistical models  which consider
 many causes of failure.  Individuals in the study  are followed from a common  initial point until the ocurrence  of the first of the events considered. Such occurrence   precludes the observation of the rest of events, acting as a censoring event.

\begin{figure}[H]
  \centering
  \includegraphics[width=3cm]{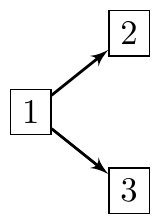}
  \caption{Competing risks model with two possible causes of failure.}\label{fig:competing}
\end{figure}

 We assume, without loss of generality,  a competing risk model with an initial state, 1, and two possible events of interest, 2 and 3 (See Figure \ref{fig:competing}). Let $T_ {1j}$ be   the time from the start point (1) to event $j=2,3$. Consider $T=\min \{T_{12}, T_{13}\}$ the time to the occurrence of the first event  and define  the cause-specific hazard function of experiencing the event $j$ in the presence
of the   competing event  as
\begin{equation}
\label{eqn:cause-specific}
h_{1j}(t)=\lim_{\Delta t \rightarrow 0}\dfrac{P(t\leq T <t+\Delta t , \delta=j \mid T  \geq t)}{\Delta t}, \quad t>0,
\end{equation}

\noindent where $\delta$ represents the indicator of the cause of the failure in such a way that
$\delta = j$ when the observation is failed from cause $j$,  and $\delta = 0$  when the
 observation is censored for events $j = 2,3$.

\noindent The cumulative   hazard function for cause $j$ at time $t$ can be defined from (\ref{eqn:cause-specific})  as $H_{1j}(t)=\int_0^t h_{1j}(u) \mbox{d}u$.
The overall survival function $S(t)$ accounts for the probability of not failing from any cause at time $t$ and it is defined as
\begin{equation}
\label{eqn:overall-survival}
S(t)=P(T>t)=\exp{\Big\{ -(H_{12}(t)+ H_{13}(t)) \Big\}}.
\end{equation}
The cumulative incidence function for cause $j$ at  $t$ (also known as crude cumulative incidence function or subdistribution function) describes the probability of failing from cause $j$ before time $t$ and it is given by
\begin{equation}
F_{1j}(t)=P(T\leq t, \delta=j)=\int_0^t h_{1j}(u)\, S(u) \,\mbox{d}u.
\end{equation}
This   probability is not a proper distribution function because  $F_{1j}(t)$ does not go to 1 when $t$ goes to $\infty$. In particular, it equals to the probability of failure from cause $j$, $F_{1j}(\infty)=P(\delta=j)$.
Cumulative incidence functions are very meaningful, as they can be interpreted as the proportion of individuals who failed by cause $j$ before time $t$. They are the main outcomes of any competing risks analysis.

Cox proportional hazard models (Cox, 1972)  introduce covariates on the cause-specific hazard functions. In particular, the cause-specific hazard of cause $j$ at time $t$ for
an individual of the target population with covariates $\boldsymbol x$ is
\begin{equation}
h_{1j}(t)=h_{0,1j}(t) \,\mbox{exp}\{ \boldsymbol{\beta}^{(1j)\prime} \boldsymbol x\},   j=2,3, \quad t>0
\end{equation}
\noindent where $h_{0,1j}(t)$ is the baseline cause-specific function of failure $j$ and $\boldsymbol{\beta}^{(1j)}$ is the vector of regression coefficients.

\subsection{Multi-state models}

\begin{figure}[H]
  \centering
  \includegraphics[width=3cm]{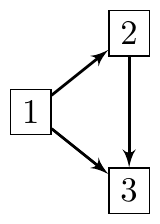}
  \caption{Illness-death model with an    initial state (1),   a transient state for illness (2), and an absorbing state such as death (3).}\label{fig:multistate}
\end{figure}

Multi-state models account for  different structures depending on the number of states and how they relate to each other (Andersen and Keiding, 2002). We focus on a particular type of multi-state model known as the  illness-death  or disability model. This is  a multi-state model with three states, which generally represent different conditions, usually health, illness and death. A healthy person can go directly to suffering from a certain illness or die without having suffered from that illness. But it is also possible that they die after having suffered from the illness. We are now faced with two competitive events but one of them is of a transitory nature and allows access to the terminal event, death (See Figure \ref{fig:multistate}). This model makes possible to study the risk of death whether the individual has been previously ill or not.

The methodological framework of multi-state models in survival analysis is initially analogous to that of competing risks. In the case of the cause-specific hazards approach we focus on the hazard function for the survival times from state 1 to 2, $T_{12}$, from 1 to 3, $T_{13}$ , and from 2 to 3, $T_{23}$ as follows:
\begin{align}
h_{1j}(t)&=\mylim{\dfrac{P(t\leq T_{1j}<t+\Delta t \mid T_{1j} \geq t)}{\Delta t}}{\Delta t \rightarrow 0}, \quad j=2,3 \nonumber \\
h_{23}(t-t_{12}| & T_{12}=t_{12}) =\mylim{\dfrac{P(t-t_{12}\leq T_{23}<t-t_{12}+\Delta t \mid T_{23} \geq t-t_{12},  T_{12}=t_{12})}{\Delta t}}{\Delta t \rightarrow 0}.
\label{eqn:hazardmulti}
\end{align}

 In many real problems (e.g. those associated with relapses of cancer diseases) the behaviour of $T_{23}$ often depends on the characteristics of transition 1 to 2. This situation is reflected in the hazard in (\ref{eqn:hazardmulti}) which depends on the time when ``illness  occur'' ($T_{12}=t_{12}$), as it represents the hazard of dying after illness.

Multi-state models are also a particular class of stochastic processes so their analysis is enriched with the procedures associated with this type of models. In this regard,  transition probabilities are conditional probabilities $p_{ij}(s,t )$ defined as the probability of being at the state $j$ at time $t$ given that the individual was in state $i$ at $s$. In the case of illness-death models, the relevant transition probabilities are $p_{11}(s,t )$, $p_{22}(s,t \mid t_{12} )$, which  assess the permanence in states 1 and 2, respectively,  $p_{12}(s,t )$ and
 $p_{13}(s,t )$  which quantify the transition to the illness and the death state from the initial state, respectively and,   $p_{23}(s,t \mid t_{12} )$ for the transition from the illness state to death. Probabilities
$p_{22}(s,t \mid t_{12} )$ and $p_{23}(s,t \mid t_{12} )$ account for the transition time $t_{12}$ from the healthy to the illness state. Note that $p_{33}(s,t \mid t_{12} )=1$ as death is a terminal state.

In practice, transition probabilities are calculated through transition intensities.  The specific linkage between transition probabilities and hazard functions is as follows  (Armero \textit{et al.}, 2016):
\begin{align}
&p_{11}(s,t)=\exp{\Big\{-\int_s^t \,(h_{12}(u)+h_{13}(u)) \mbox{d}u \Big\}}\nonumber\\
&p_{22}(s,t \mid t_{12})=\exp{\Big\{-\int_s^t h_{23}(u-t_{12}|t_{12})\mbox{d}u \Big\}} \nonumber\\
&p_{12}(s,t )=\int_s^t p_{11}(s,u) h_{12}(u)p_{22}(u,t| u) \mbox{d}u \nonumber\\
&p_{13}(s,t)=1-p_{11}(s,t)-p_{12}(s,t)\nonumber\\
&p_{23}(s,t \mid t_{12})=1-p_{22}(s,t)\nonumber\\
&p_{33}(s,t)=1,
\end{align}

It is important to mention that $p_{11}(s,t)$, the probability associated with the permanence at the initial state (1) during the time interval $[s,t]$, is also well defined for a competing risks model, with the same interpretation.

\subsection{Competing risks model as a multi-state model}

A competing risk model  can be considered as a multi-state model with  two types of events: an initial state and two states for the competing causes of failure which can be  terminal and without connection among them. The only transitions between states with probability not zero are the two that start from the initial state to each of the other two states. Transition probabilities between the states can also be expressed in terms of the subsequent transition intensities as follows (Andersen and Keiding, 2010):

\begin{align}
&p_{11}(s,t)=\exp{\Big\{-\int_s^t \,(h_{12}(u)+h_{13}(u)) \mbox{d}u \Big\}}\nonumber\\
&p_{1j}(s,t)= \int_s^t\,\exp{\Big\{-\int_s^u\, (h_{12}(x)+h_{13}(x))\, \mbox{d}x \Big\} }\, h_{1j}(u)\, \mbox{d}u  ,\,j=2,3 \nonumber \\
&p_{22}(s,t)= p_{33}(s,t)=1.
 \end{align}

 It is worth mentioning that the overall survival function $S(t)$ defined in (\ref{eqn:overall-survival}) within the competing risk scenario is the transition probability that assesses the permanence of the process in the initial state at time $t$, $S(t)=p_{11}(0, t)$. Moreover, the cumulative incidence function for cause 2 and for cause 3, $F_{12}(t)$ and $F_{13}(t)$, in the competing risk framework are the transition probabilities $p_{12}(0,t)$ and  $p_{13}(0,t)$, respectively.

\section{The PREV2FO study: Incidence of refracture and death after an osteoporotic hip fracture.}
\subsection{Problem, objective and data}

Hip fracture is a main complication of osteoporosis in elder populations. Beyond its economic cost for health care systems and society, it results in significant reductions in quality of life, disability, morbidity and mortality (Lips et al., 2005, Hopkins et al., 2016, Hernlund et al., 2013). Patients after a hip fracture are at much higher risk of a subsequent hip fracture (Johnell et al. 2004, Ryg et al. 2009) as well as death, and excess of mortality is even higher after a second hip fracture (Abrahamsen et al., 2009, Sobolev et al., 2015).

For the case study we used the PREV2FO cohort, a population‐based cohort of all patients aged 65 years and older, discharged alive after hospitalization for osteoporotic hip fracture from January 1, 2008, to December 31, 2015. Patients were followed from the date of discharge (time zero) to December 31, 2016 (end of study) or death. The study was conducted in the region of Valencia in Spain (around 5 million inhabitants, representing 10\% of the whole country population) providing free, universal health care services (besides drug cost‐sharing) to 97\% of the region's population. Data were obtained from the Valencia Health System Integrated Database (VID). The VID is the result of the linkage, by means of a single personal identification number, of a set of publicly owned population‐wide health care, clinical, and administrative electronic databases in the region of Valencia, Spain, which has provided comprehensive information since 2008 (Garcia et al. IJE., 2020).

From 34$\hspace{0.04cm}$491 patients discharged alive after hip fracture, 25$\hspace{0.04cm}$807 (74.8\%) were women and 8$\hspace{0.04cm}$684 (25.2\%) men. Regarding age, 12.4\% of patients were under 75 years old, 43.6\% were between 75 and 85 years old, 40.6\% were between 85 and 94 years old, and 3.4\% were over 95 years old. The mean age at the first fracture was 83.4 years (IQR: 79.0-88.3). Patients were followed a median time of 5.0 years (IQR: 3.0-7.0 years).

\begin{figure}[H]
  \centering
  \includegraphics[width=10cm]{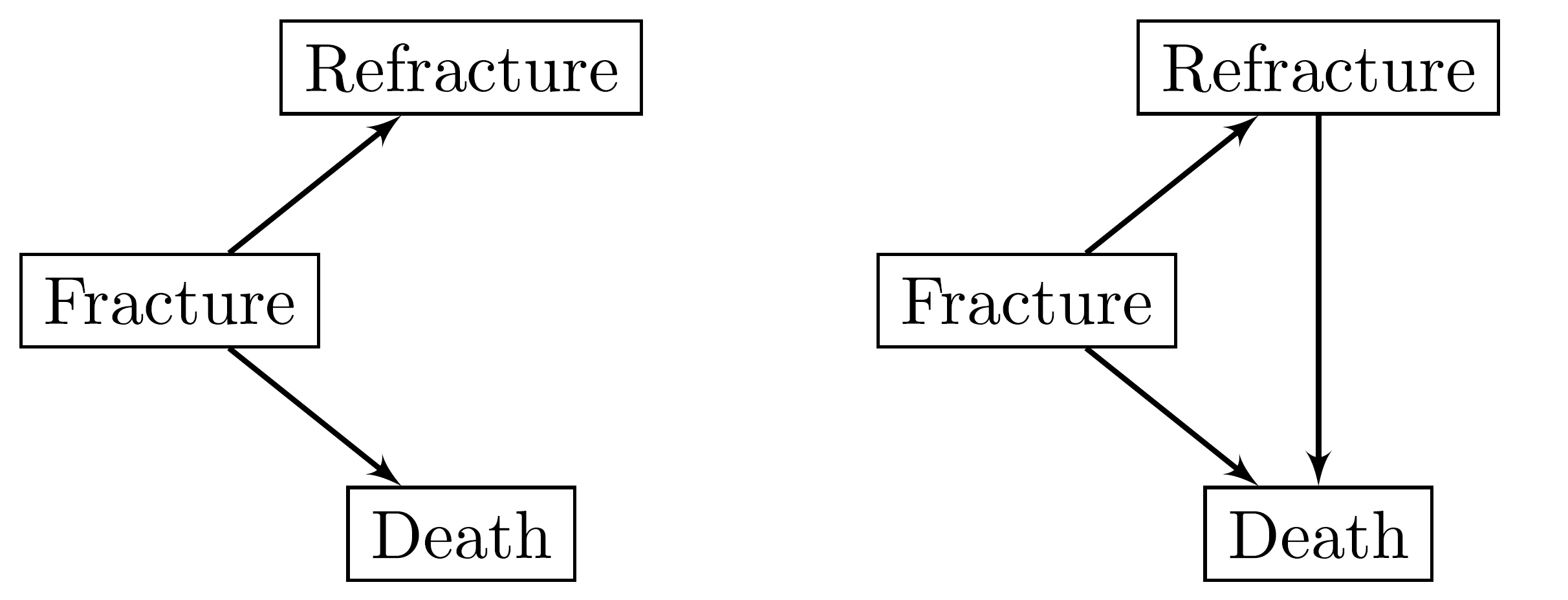}
  \caption{Competing risks model with recurrent hip fracture and death as causes of failure (left panel). Illness-death model with an initial state of hip fracture, a recurrent hip fracture state and a death state (right panel).}\label{fig:esq_prev2fo}
\end{figure}

The estimation of the incidence of recurrent hip fractures and death after an osteoporotic hip fracture, one of the main objectives of the study, could be approached by means of a competing risk model when the refracture condition is considered as a terminal event which competes with the risk of death or as an illness-death multi-state model in the case that the refracture condition can be considered as a transitory state towards death. Figure \ref{fig:esq_prev2fo} shows both approaches.

We started by working with a competing risk model with two possible causes of failure, recurrent fracture ($R$) and death ($D$) from the initial state discharge after fracture ($F$) and as relevant survival times the time from discharge after fracture to refracture, $T_{FR}$, and time from discharge after fracture to death, $T_{FD}$. Later on we discussed an illness-death model with the same three states as in the competing model but allowing the access to death from the refracture state and the subsequent time from refracture to death, $T_{RD}$ (See Figure \ref{fig:esq_prev2fo}). The latter model has a structure much closer to the reality of the problem than the one proposed by the competing risks framework. The number of patients in each transition ($F$ to $R$, $F$ to $D$, and $R$ to $D$) for women and men is shown in Figure \ref{sup:diagram} of supplementary material.

\subsection{Bayesian modeling}

Cox proportional hazards models are very popular regression models (Cox, 1972). They express the hazard function of the survival time of interest in terms of   a baseline hazard function, which determines
 the temporal shape of the hazard, and   an exponential  term which includes the relevant covariates. Cox proportional hazards models with Weibull baseline hazard functions, and gender and age at discharge as covariates have been used to model the random behaviour of times $T_{FR}$ and $T_{FD}$ in the competitive risk model and of times $T_{FR}$, $T_{FD}$, and $T_{RD}$   in the multi-state model in terms of the hazard functions  as follows:

\begin{itemize}
\item[]    \textsc{Competing  risks model:}
\begin{align*}
h_{FR}(t)=&h_{0,FR}(t) \,\mbox{exp}\{\beta_1^{(FR)}\textrm{I}_{Woman} +\beta_2^{(FR)}\textrm{Age}\},   \\
h_{FD}(t)=&h_{0,FD}(t) \,\mbox{exp}\{\beta_1^{(FD)}\textrm{I}_{Woman} +\beta_2^{(FD)}\textrm{Age}\}
\end{align*}

\noindent where $\textrm{I}_{Woman}$ is an indicator variable  that takes the value 1 when the individual is considered as a woman and zero if he is registered as a man.  $h_{0,FR}(t)$ and $h_{0,FD}(t)$ is the baseline hazard function for $T_{FR}$ and $T_{FD}$, respectively, which we define below as

\begin{equation}
h_{0,FR}(t)=\alpha^{(FR)}\lambda^{(FR)} t^{\alpha^{(FR)}-1}; \,\,h_{0,FD}(t)=\alpha^{(FD)}\lambda^{(FD)} t^{\alpha^{(FD)}-1},
\end{equation}

\noindent being  $\beta_1^{(FR)},\beta_2^{(FR)}$ ($\beta_1^{(FD)},\beta_2^{(FD)}$)   the regression coefficients associated to sex and age, respectively, for time  $T_{FR}$ ($T_{FD}$).  \\

\item[]    \textsc{Illness-death model:}

The definition of the hazard functions for the survival times in the the illness-death model were analogous to the ones for the competing risk model. The common transitions, i.e  from state $F$ to $R$ and from $F$ to $D$, were modelled identically. However, transition from refracture to death   required to be modelled with consideration to the time when patients refractured as follows
\begin{equation}
h_{RD}(t-t_{FR} \mid T_{FR}=t_{FR})=h_{0,RD}(t-t_{FR}\mid T_{FR}=t_{FR}) \,\mbox{exp}\{\beta_1^{(RD)}\textrm{I}_{Woman}+\beta_2^{(RD)}\textrm{Age}\},
\end{equation}
\noindent for $0<t_{FR}<t$, where $h_{0,FD}(t\mid T_{FR}=t_{FR})$ is the baseline hazard function and $\beta_1^{(RD)},\beta_2^{(RD)}$ are the regression coefficients associated to sex and age, respectively. The baseline hazard function was defined as
\begin{equation}
h_{0,RD}(t-t_{FR}\mid T_{FR}=t_{FR})=\alpha^{(RD)}\lambda^{(RD)} (t-t_{FR})^{\alpha^{(RD)}-1}.
\end{equation}

Note that although we modelled the common transitions in the competing risk model and the illness-death model identically, they are different models. Despite the fact that the common parameters have the same names, they are not equal because of their distinct provenance.

\end{itemize}

We used a Bayesian approach to estimate the parameters of both models. We have followed the usual steps of the Bayesian inferential process: specification of a prior distribution  for the parameters of the model, construction of the likelihood function, and computation of the posterior distribution  via Bayes' theorem.

The prior distribution is elicited assuming a prior independence and non-informative scenario.
Wide normal prior distributions were selected for the regression coefficients, $\boldsymbol{\beta}$, and Gamma distributions for the shape and scale parameters $\boldsymbol{\alpha}$ and $\boldsymbol{\lambda}$ (See Alvares \textit{et al.}, (2021) for a wide tutorial on Bayesian survival models). Posterior distributions were approximated via Markov Chain Monte Carlo methods (MCMC) as well as through the integrated nested Laplace approximation (INLA). The former are simulation procedures  very used in Bayesian statistics for dealing with non analytical posterior distributions (Chen, Shao, and Ibrahim,  2000). The latter method computes approximations of the posterior marginals for the parameters and hyperparameters in \textit{latent Gaussian models}, a subset of models which includes generalized regression models, smoothing splines models, spatial and survival models, among others (Rue, Martino and Chopin, 2009). In particular, age as a covariate is included in the model as mean-centered in order to improve the convergence of the subsequent computational Markov chain Monte Carlo methods used.  \\

\subsection{Results}

\begin{table}[h]
\centering
\begin{tabular}{c|c|cc|cc}
  && \multicolumn{2}{c|}{Competing risks model} & \multicolumn{2}{c}{Illness-death model}\\
\hline
 Transition &\textrm{Parameter} & Mean & SD  & Mean & SD  \\
  \hline
 From $F$ to $R$ & $\alpha^{(FR)}$ & 0.9197 & 0.0156   & 0.9198 & 0.0157   \\
  &$\lambda^{(FR)}$ & 0.0279 & 0.0013   & 0.0279 & 0.0013   \\
  &$\beta_1^{(FR)}$ & 0.0254 & 0.0496   & 0.0262 & 0.0486  \\
  & $\beta_2^{(FR)}$ & 0.0244 & 0.0030   & 0.0244 & 0.0030   \\
  \hline
  From $F$ to $D$&$\alpha^{(FD)}$ & 0.7759 & 0.0051    & 0.7759 & 0.0051   \\
  &$\lambda^{(FD)}$ & 0.3310 & 0.0051   & 0.3311 & 0.0050  \\
  &$\beta_1^{(FD)}$ &-0.5088 & 0.0169   & -0.5092 & 0.0166   \\
  &$\beta_2^{(FD)}$ & 0.0705 & 0.0012   & 0.0705 & 0.0012  \\

   \hline
From $R$ to $D$& $\alpha^{(RD)}$  & &&  0.6234  & 0.0154  \\
&  $\lambda^{(RD)}$ & &&  0.5769  & 0.0329   \\
&  $\beta_1^{(RD)}$ & &&  -0.6127 & 0.0655   \\
 & $\beta_2^{(RD)}$ & &&  0.0498  & 0.0046   \\
\hline
\end{tabular}
\caption{Summary of the posterior distribution of the parameters for the competing risks model and of the illness-death model.}
\label{parameters}
\end{table}

\subsubsection*{Posterior distribution}
The approximate posterior distribution of the competing risk and of the illness-death model parameters have been summarized in Table \ref{parameters} via their posterior mean and standard deviation. No differences have been observed between the results of the competing risk and the illness-death model for the two common transitions, from fracture to refracture and from fracture to death. However, the illness-death model provides additional information regarding the transition from refracture to death.

\subsubsection*{Refracture and death}

The cumulative incidence function for refracture (death) assesses the probability that a person refractures (dies) before a certain time. These probabilities depend on the parameters of the model and consequently their posterior distribution is determined by the posterior distribution computed previously.
Competing risk and illness-death models provide almost identical results for the incidence of refracture and death without refracture (See Table \ref{tab:incid}, posterior mean of the one-year cumulative incidence of refracture and death from any cause by sex and age). However, the illness-death model provides additional information regarding the probability of transition from refracture to death (or the cumulative incidence of death after refracture).

\begin{table}[h]
\centering
\begin{tabular}{c|c|c|cc|cc}
 &&& \multicolumn{2}{c|}{Competing risks model} & \multicolumn{2}{c}{Illness-death model}\\
\hline
 Transition & Sex & Age & Mean & 95\% CI  & Mean & 95\% CI  \\

  \hline

  \multirow{6}{*}{From $F$ to $R$}& Women & 70  & 1.96 & (1.76, 2.16) & 1.96 & (1.76, 2.16)  \\
  &&80 &  2.39 & (2.23,  2.55) & 2.39 & (2.24,  2.55)  \\
  &&90 &  2.80 & (2.61,  2.99) & 2.80 & (2.60,  3.00)   \\
 \cline{2-7}
 &Men & 70  & 1.86 & (1.65,  2.08) & 1.86 & (1.65,  2.09)     \\
      &&80  & 2.21 & (2.01,  2.42) & 2.21 & (2.01,  2.42)  \\
      &&90  & 2.46 & (2.21,  2.72) & 2.45 & (2.21,  2.72)   \\

  \hline
 \multirow{6}{*}{From $F$ to $D$}&Women & 70  &    7.36 & (7.07,  7.68) & 7.36 & (7.05,  7.67)   \\
  &&80 &  14.30 & (13.96,  14.65) & 14.30 & (13.95,  14.63)   \\
  &&90 &  26.73 & (26.20,  27.27) & 26.72 & (26.17,  27.26)   \\
  \cline{2-7}
& Men & 70  &   11.94 & (11.43,  12.46) & 11.95 & (11.43,  12.46 )   \\
  &&80 &  22.63 & (21.99,  23.28) & 22.63 & (22.02,  23.29)   \\
  &&90 &  40.34 & (39.33,  41.36) & 40.35 & (39.42,  41.36)  \\
  \hline

 \multirow{6}{*}{From $R$ to $D$}&Women & 70  &    &  & 14.77 & (12.73,  16.87)  \\
 & &80 &  &    & 23.11 & (21.52,  24.79)   \\
  &&90 &  &   & 35.14 & (32.75,  37.60)   \\
  \cline{2-7}
 &Men & 70  &    &  & 25.56 & (22.01,  29.37)    \\
  &&80 &  &    & 38.46 & (35.25,  41.87)   \\
  &&90 &  &   & 55.03 & (50.61,  59.59)   \\
  \hline

\end{tabular}
\caption{Posterior mean and 0.95 credibility interval of the one-year cumulative incidences of refracture, death without refracture and death after refracture for the competing risks model and for the illness-death model, by sex and age.}
\label{tab:incid}
\end{table}

Overall, refracture probability increases with age and women have also higher incidence (one-year incidences of 1.96\% at 70 and 2.80\% at 90 years old for women, whilst 1.86\% and 2.46\% for men). Regarding death without refracture, incidence increases with age and has been estimated higher for men (one-year incidences of 7.36\% at 70 and 26.73\% at 90 years old among women, \emph{versus} 11.94\% and 40.34\% for men).

\subsubsection*{Death after refracture}

\begin{figure}[h]
  \centering
  \includegraphics[width=17cm]{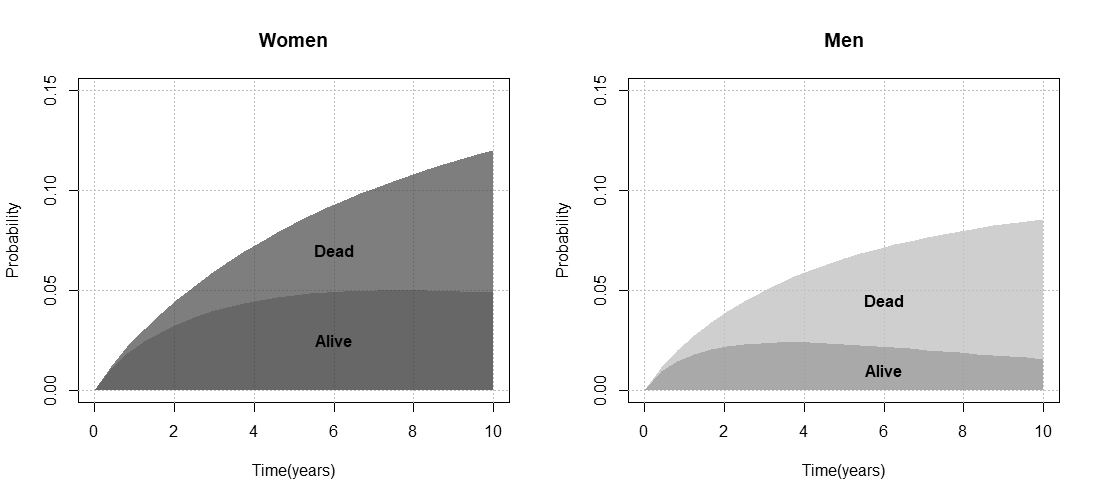}
  \caption{Posterior mean of the cumulative incidence of refracture (upper curve) and of the transition probability from initial state to refracture state (lower curve). Light area in both figures represents the proportion of dead patients who had previously suffered a refracture while the dark area represents the proportion of refractured people still alive.}\label{fig:prob_FR}
\end{figure}

The illness-death model provides relevant information about the transition from refracture to death that the competing risks model cannot address. Figure \ref{fig:prob_FR} shows the posterior mean of the cumulative incidence of refracture and the posterior mean of the transition probability from the initial state of fracture to the refracture state. Both outcomes are defined in relation to a specific time $t$. The upper curve corresponds to the total accumulation of refractures (cumulative incidence) occurred before time $t$ and the lower curve refers to the probability that a patient is in the state of refracture at time $t$, and therefore alive. Thus, it should be understood that at time $t$ the dark shaded area shows living patients who have refractured and the light shaded area shows refractured patients who have died.

In broad terms, death after refracture shows a similar pattern to death without refracture, i.e incidence increases with age and is higher for men (one-year incidences of 14.77\% at 70 and 35.14\% at 90 years old for women, whereas 25.56\% and 55.03\% for men).

\section{Concluding remarks}

Several authors have addressed the relation between competing risks and multi-state models, with consensus on identifying the first as
a particularization of the second (Andersen, 2002; Putter, 2007). Moreover, some authors performed comparisons between
both model estimations (Leffondré, 2013). However, there is no literature available
comparing both estimations from a Bayesian perspective. In this paper, we showed that not only can the competing risks model be considered
as a particular case of an illness-death model, but also they provide identical information of the common transitions under some given conditions.
It is a matter of prior independence of the parameters and separability of the likelihood. Regarding likelihood from both models, it presents a separated structure, with separated products for each transition. It results in a sort of ``independent" transitions,
what makes the inclusion of a new transition after refracture not to change the estimations of the other transitions. Had transitions had shared parameters, estimations might have differed. Correlated random effects would be also supposed to violate the separability, resulting in different estimations from a competing risks and an illness-death model. Finally, non-informative and independent priors were assumed for the parameters. Prior correlation between parameters might translate into posterior correlation and thus posterior distributions would be sensitive to the addition of new transitions.

An important consideration should be made about the estimation of the incidence of death among the population. Although the incidence of refracture can be estimated equally from both of our models, it is not the case with the incidence of death. The proposed competing risks model only accounts for death without refracture, as the observation of death is censored for those refractured patients. However, when including death after refracture through the illness-death model, the total incidence of death becomes a natural outcome. In particular, it is defined as the transition probability from the initial state at the beginning of the follow-up time to death state. This advantage, always under the aforementioned assumptions and considerations, makes the illness-death model a preferable tool/method to assess cumulative incidences when studying more than one event or when death is an unavoidable event to consider (e.g. elder population).

\section*{Funding}
The PREV2FO cohort study was funded by Spanish Ministry of Science, Innovation, and Universities (grants PI14/00993 and PI18/01675, cofinanced by the European Regional Development Fund), and awarded with the Merck Research Grant 2019 in health outcomes and research from Merck Salud Foundation. F.LL. was funded by the Instituto de Salud Carlos III, Spanish Ministry of Science, Innovation, and Universities, cofinanced by the European Social Fund (Pre-doctoral grant Pfis no. FI19/00190).
C.A.  was partially funded by PID2019-106341GB-I00 from the Ministerio de Ciencia e Innovación (MCI, Spain).

\section*{Ethics}

The study was approved by the Ethics Committee for Clinical Research of the General Directorate of Public Health and the Centre for Public Health Research (session on October 26, 2012). All protocols were performed in accord with Spanish laws on data protection for health research (Act 3/2018 transposing the 2015 European Data Protection Regulation).

\section*{References}

\begin{itemize}[leftmargin=*]
\item[]	Abrahamsen B, van Staa T, Ariely R, Olson M, Cooper C. Excess mortality following hip fracture: a systematic epidemiological review. \emph{Osteoporos Int.} 2009;\textbf{20}(10):1633–1650.

\item[] Alvares D,  Lázaro E,  Gómez-Rubio V, and Armero C. Bayesian survival analysis with BUGS. \emph{Statistics in Medicine} 2021;1–46. doi:10.1002/sim.8933.

\item[] Andersen PK and Keiding N. Multi-State Models for Event History Analysis. \textit{Statistical Methods in Medical Research} 2002;\textbf{11}(2):91-115.

\item[] Andersen PK, Abildstrøm SZ, Rosthøj S. Competing risks as a multi-state model. \emph{Statistical Methods in Medical
Research} 2002;\textbf{11}:203–215.

\item[] Andersen K, Geskus RB, de Witte T, and Putter H. Competing risks in epidemiology: possibilities and pitfalls
 \textit{International Journal of Epidemiology} 2012;\textbf{41}:861–870.

 \item[] Armero C, Cabras S, Castellanos ME, and Quirós A. Bayesian Analysis of a Disability Model for Lung Cancer Survival. \textit{Statistics Methods in Medical Research} 2016;\textbf{25}(1):336–351.

\item[] Beyersmann J, Wolkewitz M, Allignol A, Grambauer N, Schumacher M. Application of multistate models in hospital epidemiology: advances and challenges
\textit{Biometrical Journal} 2011; \textbf{53}(2):332-50.

\item[] Chen M-H, de Castro M, Ge M, and Zhang Y. Bayesian Regression Models for Competing Risks.  In J. P. Klein,
H. C. van Houwelingen, J. G. Ibrahim, and T. H. Scheike, editors, \textit{Handbook of Survival Analysis}: chapter
2, pages 180-198. Chapman \& Hall/CRC, Boca Raton; 2014.

\item[] Chen M-H, Shao Q-M, and Ibrahim JG. \textit{Monte Carlo Methods in Bayesian Computation.}, Springer, New York; 2000.

\item [] Commenges D. Multi-state Models in Epidemiology. \textit{Lifetime Data Analysis} 1999;\textbf{5}:315–327.

\item[] Cox  DR.  Regression models and life-tables. \textit{Journal of the Royal
Statistical Society, Series B} 1972;\textbf{34}:87–220.

\item[] Gail M. A review and critique of some models used in competing risk analyses. \emph{Biometrics} 1975;\textbf{31}:209–222.

\item[]	García-Sempere A, Orrico-Sánchez A, Muñoz-Quiles C, Hurtado I, Peiró S, Sanfélix-Gimeno G, Diez-Domingo J. Data Resource Profile: The Valencia Health System Integrated Database (VID). \emph{Int J Epidemiol.} 2020;\textbf{49}(3):740-741e. doi: 10.1093/ije/dyz266.

\item[] Gaynor JJ, Feuer EJ, and Tan CC. On the use of cause-specific failure and conditional failure probabilities: examples from clinical oncology data. \emph{Journal of the American Statistical Association} 1993;\textbf{88}:400–409.

\item[] Hernlund E, Svedbom A, Ivergård M, Compston J, Cooper C, Stenmark J, et al. Osteoporosis in the European Union: medical management, epidemiology and economic burden. \emph{Arch. Osteoporos} 2013;\textbf{8}:136. doi: 10.1007/s11657-013-0136-1.

\item[] Hill M, Lambert PC, and  Crowther MJ.
Relaxing the assumption of constant transition rates in a multi-state model in hospital epidemiology. \textit{BMC Medical Research Methodology} 2021;\textbf{21}:16.

\item[] Hopkins RB, Burke N, Von Keyserlingk C, et al. The current economic burden of illness of osteoporosis in Canada. \textit{Osteoporos Int.} 2016;\textbf{27}(10):3023–3032.

\item[] Johnell O, Kanis JA, Oden A, et al. Fracture risk following an osteoporotic fracture. \emph{Osteoporos Int.} 2004; \textbf{15}(3):175–9.

\item[] Lau B, Cole SR, and Gange SJ. Competing Risk Regression Models for Epidemiologic Data.
\textit{American Journal of Epidemiology} 2009; 170(2):244–256.

\item[] Leffondré K, Touraine C, Helmer C, Joly P. Interval-censored time-to-event and competing risk with death: is the illness-death model more accurate than the Cox model? \emph{International Journal of Epidemiology} 2013;\textbf{42}(4):1177–1186. doi:10.1093/ije/dyt126.

\item[] Lips P, van Schoor, NM. Quality of life in patients with osteoporosis. \textit{Osteoporos Int.} 2005; \textbf{16}(5):447–455.

\item[] Putter H, Fiocco M, and Geskus R. Tutorial in biostatistics: competing risks and
multi-state models. \emph{Statistics in Medicine} 2007;\textbf{26}(11):2277–2432.

\item[] Rue H, Martino S, Chopin N. Approximate Bayesian inference for latent Gaussian models by using integrated nested Laplace approximations. \emph{Journal of the Royal Statistical Society: Series B (Statistical Methodology)} 2009;\textbf{71}:319-392. doi:10.1111/j.1467-9868.2008.00700.x.

\item[]	Ryg J, Rejnmark L, Overgaard S, Brixen K, and Vestergaard P. Hip fracture patients at risk of second hip fracture: a nationwide population-based cohort study of 169,145 cases during 1977-2001. \emph{Journal of bone and mineral research: the official journal of the American Society for Bone and Mineral Research} 2009;\textbf{24}(7):1299–1307. doi:10.1359/jbmr.090207.

\item[] Schuster  NA, Hoogendijk  E  O, Kok  AAL,
 Twiska JWR, and Heymans MW. Ignoring competing events in the analysis of survival data may lead to
biased results: a nonmathematical illustration of competing risk analysis. \textit{Journal of Clinical Epidemiology} 2020;\textbf{122}:42-48.

\item[]	Sobolev B, Sheehan KJ, Kuramoto L, Guy P. Excess mortality associated with second hip fracture. \emph{Osteoporos Int.} 2015;\textbf{26}(7):1903-1910. doi:10.1007/s00198-015-3104-3.

\item[] Wei S, Tian J, Song X,  Wu B and Liu L.  Causes of death and competing risk analysis of the associated
factors for non-small cell lung cancer using the Surveillance,
Epidemiology, and End Results database. \textit{Journal of Cancer Research and Clinical Oncology} 2018;\textbf{144}:145–155.

\end{itemize}

\section*{Supplementary}

\begin{figure}[H]
  \includegraphics[width=16cm]{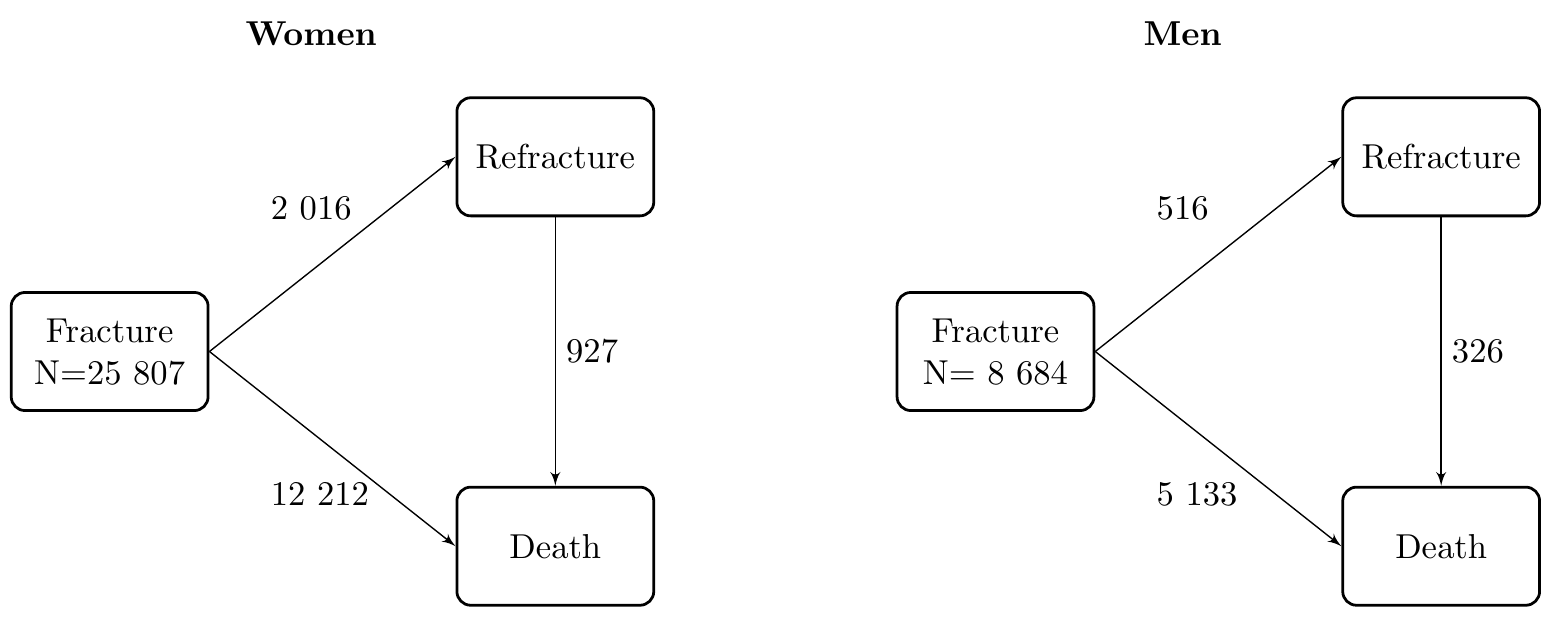}
  \caption{Number of patients in each transition by sex.}\label{sup:diagram}
\end{figure}

\end{document}